\pdfoutput=1
\documentclass{optica-article}

\journal{opticajournal}

\articletype{Research Article}

\usepackage{bm}  
\usepackage{braket}

\newcommand{\dbeta}{\Delta\beta}
\newcommand{\pp}{\psi_{+}}
\newcommand{\pmm}{\psi_{-}}
\newcommand{\xo}{x_{0}}
\newcommand{\sy}{\sigma_{y}}

\begin{document}

\title{Transverse momentum as the counter-diabatic generator in bent waveguide couplers}

\author{Yu-Kai Lee\authormark{1} and Shuo-Yen Tseng\authormark{1,*}}

\address{\authormark{1}Department of Photonics, National Cheng Kung University, Tainan City 701, Taiwan}

\email{\authormark{*}tsengsy@mail.ncku.edu.tw}

\begin{abstract*}
Axis bending of a mode-evolution coupler realizes the counter-diabatic (CD)
protocol [Opt. Express \textbf{34}, 1214 (2026)]. We identify the operator
responsible. Rigid lateral displacement is generated by the transverse momentum
$\hat p_x$, which in a real two-mode supermode basis is purely $\sy$: exactly
the form of the CD term, and the one form that detuning and coupling cannot
supply. Equating the two gives the axis slope in closed form, as a ratio of two
matrix elements at a single cross section, or
$\dot{\xo}=\dot\theta/(\gamma\dbeta^{2})$ in coupled-mode variables. It agrees
with a numerical offset search to $0.3\%$ and is shown to be the earlier
curvature condition viewed from the Kramers--Henneberger frame. Beam
propagation simulations compare the bent coupler directly with the
unitary-transformation realization built from the same reference structure:
the two reach supermode fidelities of $0.994$ and $1.000$ against $0.886$ for
the untreated coupler, show comparable bandwidth and fabrication tolerance, and differ
inside the device by the frame transformation that relates them.
\end{abstract*}

\section{Introduction}

Light propagation in coupled waveguides is governed, under the scalar and
paraxial approximations, by coupled-mode equations that are formally identical
to the time-dependent Schr\"odinger equation of a few-level quantum system
\cite{Dragoman2004,Longhi2009}. This analogy has allowed adiabatic passage
techniques developed in atomic physics \cite{Bergmann1998} to be carried over to
integrated optics. The resulting mode-evolution couplers transfer light by the
slow evolution of a single supermode, and because their operation does not
depend on a precise beat length they offer large bandwidth and fabrication
tolerance \cite{Syms1992,Xing2013,Wang2018}; the same principle underlies
mode-evolution polarization splitters, rotators and splitter--rotators
\cite{Watts2005a,Watts2005b,Hung2024}. Their drawback is length: the adiabatic
condition requires the geometry to vary slowly compared with the supermode beat
length, and practical devices are millimeters long. Minimizing the unwanted
mode coupling \cite{Louisell1955,Sun2009} and optimizing the taper shape
\cite{Ramadan1998,Syahriar1998,Ng2019} shorten them, but within the bound that
the adiabatic condition itself sets.

Shortcuts to adiabaticity (STA) \cite{GueryOdelin2019} were developed to obtain
the results of adiabatic evolution in a shorter distance, and have found many
applications in waveguide devices
\cite{Lin2012,Tseng2012,Tseng2014,Stefanatos2014,MartinezGaraot2017,Chung2019,Wu2020}.
Among the STA protocols, counter-diabatic (CD) driving
\cite{Unanyan1997,Demirplak2003,Berry2009,Chen2010} is the most direct: a term is added to the
Hamiltonian that cancels the nonadiabatic coupling exactly, so that the system
follows the instantaneous eigenstate at any speed. For a two-waveguide coupler
with $H_0=\Delta\sigma_z+\Omega\sigma_x$, however, the CD term is proportional
to $\sy$, whereas the two parameters that can be controlled by the geometry,
the detuning $\Delta$ through the waveguide widths and the coupling $\Omega$
through the waveguide spacing, both lie in the $\sigma_z$--$\sigma_x$ plane.
The CD term therefore cannot be implemented directly in a coupled-waveguide
system.

The established solution is a unitary transformation. Ib\'a\~nez \textit{et
al.} showed that a Schr\"odinger equation can be transformed into a family of
dynamical equations in different interaction pictures that describe the same
physical process, so that an unrealizable Hamiltonian can be replaced by a
realizable one \cite{Ibanez2012}. For coupled waveguides this route was
identified in Ref.~\cite{Tseng2013}, where the CD protocol was mapped by a
rotation about the propagation axis onto coupled-mode equations of the
original form with modified coupling and detuning, and a universal design
formalism for short mode-evolution devices was obtained; related waveguide
implementations followed \cite{Paul2015}, and the equivalence of this approach
with the other principal STA techniques was established in
Ref.~\cite{Tseng2014}. We refer to the device produced in this way as the gauge
realization. Figure~\ref{fig:device} shows the untreated coupler in panel (a),
the gauge realization in panel (b), and the bent realization discussed below
in panel (c). In the gauge realization both modified parameters must be written
into the geometry, $\Omega_s$ through the waveguide spacing and $\Delta_s$
through the width difference. Neither dimension is
easily controlled in fabrication, and the spacing is the more sensitive of the
two because the coupling depends on it exponentially \cite{Okamoto2006}.

A less commonly explored design parameter is the coupler axis. Longhi
demonstrated Landau--Zener dynamics in a directional coupler with a cubic axis
bend \cite{Longhi2005}. Bent directional couplers have since been used to
desensitize the coupling to wavelength and fabrication
\cite{Doerr2005,Morino2014}, axis bending of uniform-width couplers has yielded
compact and robust devices through analogies with phase-jump models
\cite{Tseng2021,Wu2022}, and an STA design of a bent mode-evolution coupler was
reported in Ref.~\cite{Chung2023}. In matter-wave optics, the curvature of a
guide has likewise been used as a control beyond the adiabatic regime
\cite{Impens2020}. Fargas Cabanillas and Popovi\'c showed that tilting
the waveguide geometry provides a degree of freedom that nulls the intermodal
coupling, determining the tilt by scanning all angles at each cross section
\cite{Cabanillas2018}. Siriani and Tambasco generalized this into a design
toolbox based on fully vectorial coupled local-mode theory, deriving a
cancellation condition between the rates of change of two geometry parameters
and applying it with the lateral position of the structure center as one of
them \cite{Siriani2021}. In both cases the cancellation is located numerically,
by perturbing the geometry and re-solving for the modes. The physical
interpretation of these bent structures was established in
Ref.~\cite{Lee2026}, which showed that axis bending realizes the CD protocol
and derived an analytical condition for the bending profile from the
coupled-mode Hamiltonian of the bent coupler in the Kramers--Henneberger frame,
in which the bend appears as a term proportional to the axis curvature.

In this work we identify the operator that makes axis bending a CD protocol.
Rather than transforming the coupled-mode Hamiltonian and matching it to the CD
Hamiltonian, we ask how a lateral displacement of the structure enters the
generator of the dynamics. The answer is fixed by an elementary fact: rigid
translation is generated by the transverse momentum $\hat p_x$, for every
structure, and in a real two-mode basis $\hat p_x$ is purely $\sy$. Among the
geometric parameters of a coupler, the axis position is thus the only one whose
effect on the Hamiltonian is known analytically without reference to the
particular cross section, and it points in precisely the direction that the CD
term requires and that widths and spacing cannot reach. The transverse momentum
is the CD generator of the bent coupler. The bending profile then follows in
closed form as a ratio of two matrix elements evaluated at a single cross
section of the unshifted structure, without trial geometries or numerical
optimization. The condition is first order in the axis position, and we show
that it is the curvature condition of Ref.~\cite{Lee2026} expressed in a
different frame: the two derivations contain the same $\sy$ content, carried in
the comoving frame by the boost term and in the Kramers--Henneberger frame by
the curvature term together with the transformation back to the waveguide
basis. Compared with Ref.~\cite{Lee2026}, we provide in this work the physical
interpretation of the bend through $\hat p_x$ and a design rule that gives the
axis slope directly at each cross section. The design rule does not require
solving a differential equation for the curvature. We also perform a direct comparison of the bent
realization with the gauge realization, which was not made in
Ref.~\cite{Lee2026}.

The paper is organized as follows. Section~\ref{sec:theory} presents the
derivation, its expression in coupled-mode variables, and its relation to the
curvature condition. Section~\ref{sec:verify}
verifies the closed form against the numerical offset search and by beam
propagation simulation, and then addresses a question left open by the
identification in Ref.~\cite{Lee2026}: if axis bending and the unitary
transformation are two realizations of one protocol, the two devices should
agree in the transfer evaluated at the facets and differ inside the device by
the frame transformation that relates them. We test both statements by
constructing the two devices from a common reference coupler, and we compare
their performance against device length.
Section~\ref{sec:disc} discusses the limits of the present treatment.

\section{Theory}
\label{sec:theory}

\subsection{Propagation as a Schr\"odinger equation}

Scalar paraxial propagation with reference index $n_s$,
\begin{equation}
2ik_0n_s\,\partial_z E=-\partial_x^2E-k_0^2\left(n^2-n_s^2\right)E,
\end{equation}
may be written $i\,\partial_zE=\mathcal{H}E$ with
\begin{equation}
\mathcal{H}(z)=-\frac{1}{2m}\partial_x^2+V(x,z),\qquad
m=k_0n_s,\qquad
V=-\frac{k_0}{2n_s}\left(n^2-n_s^2\right),
\label{eq:H}
\end{equation}
so that the propagation coordinate plays the role of time and $m=k_0n_s$ acts as
an effective mass. Throughout, $\Braket{f|g}=\int f^{*}g\,dx$ denotes the
transverse overlap and an overdot denotes $d/dz$.

\begin{figure}[htbp]
\centering\includegraphics[width=\linewidth]{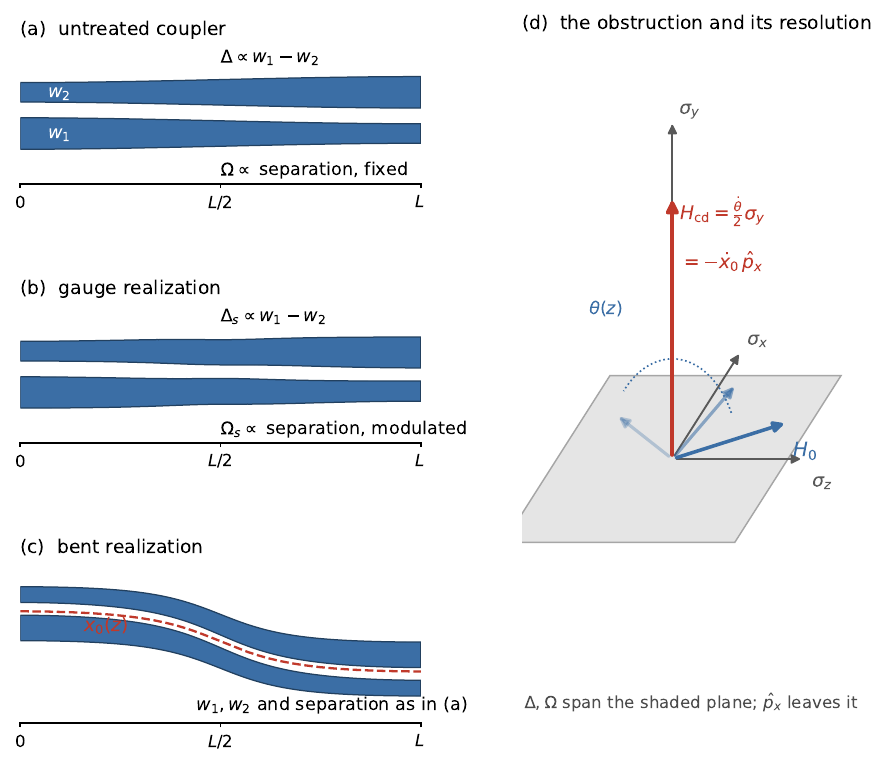}
\caption{\label{fig:device}Schematic of the coupler geometries, drawn to scale
from the facet-matched designs of Sec.~\ref{sec:equiv} ($L=1300~\mu$m). (a) Untreated mode-evolution coupler:
the widths $w_1(z)$ and $w_2(z)$ are tapered to sweep the detuning
$\Delta\propto w_1-w_2$, at fixed center-to-center spacing $d$, which sets the
coupling $\Omega$. (b) Gauge realization: the axis is straight, and both the
width difference and the spacing are modulated to write $\Delta_s$ and
$\Omega_s$ of Eq.~\eqref{eq:gauge}. (c) Bent realization: $w_1$, $w_2$ and $d$
are those of (a), and the axis is displaced along $\xo(z)$ (dashed) according to
Eq.~\eqref{eq:slope}. (d) Operator picture. $\Delta$ and $\Omega$ span the
shaded $\sigma_z$--$\sigma_x$ plane, in which $H_0$ rotates as the mixing angle
$\theta(z)$ sweeps; the CD term of Eq.~\eqref{eq:Hcd} lies along $\sigma_y$,
outside the plane. Lateral displacement supplies it, because its generator
$\hat p_x$ is purely $\sigma_y$ in a real two-mode basis, Eq.~\eqref{eq:boost}.}
\end{figure}

\subsection{Two-mode reduction and the two bases}
\label{sec:twomode}

We consider two weakly coupled waveguides of widths $w_1(z)$ and $w_2(z)$ at
center-to-center spacing $d$, as shown in Fig.~\ref{fig:device}(a). The
structure supports two guided supermodes, and we work in the two-dimensional
space they span. It is convenient to keep two bases in view.
The \emph{local} basis $\{u_1,u_2\}$ consists of the modes the two waveguides
would carry in isolation, in which the coupled-mode Hamiltonian takes the
standard form \cite{Yariv1973,Okamoto2006}
\begin{equation}
H_0(z)=\Delta(z)\,\sigma_z+\Omega(z)\,\sigma_x ,
\label{eq:H0}
\end{equation}
where $2\Delta$ is the difference of the isolated propagation constants, set by
$w_1-w_2$, and $\Omega$ the evanescent coupling, set by $d$. In a
mode-evolution coupler the widths are tapered so that $\Delta$ sweeps from a
large positive value through zero to a large negative one, while $\Omega$
remains comparatively small.

The \emph{adiabatic} basis $\{\pp,\pmm\}$ consists of the instantaneous
eigenvectors of \eqref{eq:H0}, the supermodes, which we take real and
orthonormal:
\begin{equation}
\pp=\cos\tfrac{\theta}{2}\,u_1+\sin\tfrac{\theta}{2}\,u_2,\qquad
\pmm=-\sin\tfrac{\theta}{2}\,u_1+\cos\tfrac{\theta}{2}\,u_2,
\label{eq:supermodes}
\end{equation}
with the mixing angle and the supermode splitting given by
\begin{equation}
\theta(z)=\arctan\frac{\Omega}{\Delta},
\qquad
\dbeta(z)=2\sqrt{\Delta^2+\Omega^2}.
\label{eq:theta}
\end{equation}
Equation~\eqref{eq:supermodes} states that the two bases are related by the
rotation $U_a=\exp(-i\theta\sigma_y/2)$, generated about the $y$ axis. Two
consequences will be used repeatedly. First, $H_0$ is diagonal in the adiabatic
basis, equal to $(\dbeta/2)\sigma_z$ there. Second, because $U_a$ is generated
by $\sigma_y$ itself, $\sigma_y$ has the same matrix in both bases, so a term
identified as $\sigma_y$ in one basis is $\sigma_y$ in the other and no
ambiguity arises below. The transformation between the bases mixes $\sigma_z$
and $\sigma_x$ but leaves $\sy$ alone.

Adiabatic following means that a field launched in $\pp$ remains in $\pp$. Since
the taper carries $\theta$ from near $0$ to near $\pi$, the supermode $\pp$
starts localized in one waveguide and ends localized in the other, and adiabatic
following transfers the light. Whether it does so depends on the coupling
between supermodes induced by the $z$ dependence of the structure, to which we
now turn.

\subsection{The diabatic drive is a $\sy$ term}

Let $\psi_a(x;z)$, $a=\pm$, denote the two instantaneous supermodes and expand
the propagating field as $E(x,z)=\sum_a c_a(z)\psi_a(x;z)$. Substituting into
$i\partial_zE=\mathcal{H}E$ and taking the overlap with $\psi_b$ gives
\begin{equation}
i\,\dot c_b=\mu_b c_b-i\sum_a\Braket{\psi_b|\partial_z\psi_a}c_a ,
\label{eq:cme}
\end{equation}
where $\mu_\pm$ are the eigenvalues of $\mathcal{H}$ and the second term
collects the effect of the modes themselves changing with $z$. Normalization,
$\Braket{\psi_a|\psi_a}=1$ for all $z$, forces
$\Braket{\psi_a|\partial_z\psi_a}=0$, so the matrix
$\Braket{\psi_b|\partial_z\psi_a}$ is real and antisymmetric. For two modes it
therefore has a single independent entry,
\begin{equation}
\mathcal{A}(z)\equiv\Braket{\pp|\partial_z\pmm},
\end{equation}
and, dropping an irrelevant trace, Eq.~\eqref{eq:cme} is generated in the
adiabatic basis by
\begin{equation}
H_{\mathrm{ad}}=\frac{\dbeta}{2}\,\sigma_z+\mathcal{A}\,\sy .
\label{eq:Had}
\end{equation}
The structure of \eqref{eq:Had} is the starting point for what follows: a
real antisymmetric generator expressed in a real basis is necessarily $\sy$. Differentiating \eqref{eq:supermodes} gives
$\partial_z\pmm=-(\dot\theta/2)\pp$, so that
\begin{equation}
\mathcal{A}=-\dot\theta/2 .
\label{eq:Aval}
\end{equation}
The device is adiabatic when $|\mathcal{A}|\ll\dbeta$, that is when the mixing
angle turns slowly compared with the supermode beat rate. Shortening the device
at fixed taper shape increases $\dot\theta$ and violates this condition, which
is the origin of the length penalty.

\subsection{Counter-diabatic driving and its obstruction}
\label{sec:cd}

Counter-diabatic driving removes the second term of \eqref{eq:Had} outright by
adding to the Hamiltonian the term that cancels it
\cite{Demirplak2003,Berry2009},
\begin{equation}
H_{\mathrm{cd}}=-\mathcal{A}\,\sy=\frac{\dot\theta}{2}\,\sy ,
\label{eq:Hcd}
\end{equation}
after which $|c_\pm|^2$ are constants of the motion and adiabatic following is
exact at any device length. By the remark following \eqref{eq:theta}, the form
\eqref{eq:Hcd} is the same in the local basis, so the requirement is
unambiguous: the coupler must supply a $\sy$ term of strength $\dot\theta/2$.

This is precisely what the cross section of a directional coupler cannot
provide. The two quantities set by the geometry appear in \eqref{eq:H0} as
$\Delta$, controlled by the width difference, and $\Omega$, controlled by the
spacing; both lie in the $\sigma_z$--$\sigma_x$ plane, as illustrated in
Fig.~\ref{fig:device}(d). Adding to either changes $\theta$, and hence changes $\mathcal{A}$
itself, but cannot cancel it, because the added term is diagonal in the basis in
which $\mathcal{A}$ is off diagonal. No choice of widths and gaps produces
\eqref{eq:Hcd}.

The established response is to change the picture rather than the Hamiltonian.
Ib\'a\~nez \textit{et al.} showed that a given physical process is described by
a family of dynamical equations in different interaction pictures, so that an
unrealizable Hamiltonian may be traded for a realizable member of the family
\cite{Ibanez2012}. Applying the $z$-dependent rotation about the propagation
axis,
\begin{equation}
U_z=\mathrm{diag}\!\left(e^{-i\varphi/2},\,e^{+i\varphi/2}\right),
\qquad
\varphi=\arctan\frac{\dot\theta}{2\Omega},
\label{eq:Uz}
\end{equation}
to $H_0+H_{\mathrm{cd}}$ yields a Hamiltonian of the original form
\eqref{eq:H0}, with modified parameters
\begin{equation}
\Omega_s=\sqrt{\Omega^2+\left(\dot\theta/2\right)^2},
\qquad
\Delta_s=\Delta-\dot\varphi/2 ,
\label{eq:gauge}
\end{equation}
which is realizable, since both are again a coupling and a detuning. This is the
route identified for coupled waveguides in Ref.~\cite{Tseng2013}, and we shall
call the device it produces the \emph{gauge realization}. Two features of it
matter below. Because $U_z$ is diagonal in the local basis it leaves the
waveguide populations unchanged, so the gauge device reproduces the
counter-diabatic dynamics exactly in that basis, provided $U_z$ reduces to the
identity at both facets; by \eqref{eq:Uz} this requires $\dot\theta\to0$ there.
And because $\Omega_s>\Omega$ wherever $\dot\theta\neq0$ while
$\Delta_s$ differs from $\Delta$ by $\dot\varphi/2$, \emph{both} geometric
parameters must now be modulated along the device: the separation to write
$\Omega_s$ and the width difference to write $\Delta_s$. Comparing
Fig.~\ref{fig:device}(a) with Fig.~\ref{fig:device}(b) shows the cost. The
untreated coupler tapers its widths at fixed separation; the gauge realization
must taper a different width profile \emph{and} modulate the separation, and the
latter is the less well controlled dimension because the coupling depends on it
exponentially.

\subsection{Transverse momentum as the counter-diabatic generator}
\label{sec:boost}

We now show that the required $\sy$ term is available from a degree of freedom
that \eqref{eq:H0} does not contain, because it is not a property of the cross
section: the position of the coupler axis, Fig.~\ref{fig:device}(c).

Let the whole structure be displaced laterally, $V(x,z)\to V(x-\xo(z),z)$, and
work in the comoving coordinate $x'=x-\xo$. The frame change adds the usual
boost term,
\begin{equation}
\mathcal{H}\longrightarrow\mathcal{H}(x')-\dot{\xo}\,\hat p_x,
\qquad\hat p_x=-i\partial_x .
\label{eq:boostH}
\end{equation}
Its matrix elements in the real supermode basis are strongly constrained. The
diagonal vanishes identically,
\begin{equation}
\Braket{\psi_a|\hat p_x\psi_a}=-i\int\psi_a\,\partial_x\psi_a\,dx
=-\tfrac{i}{2}\int\partial_x\!\left(\psi_a^2\right)dx=0,
\end{equation}
because a real bound mode carries no net transverse momentum, while the
off-diagonal element
\begin{equation}
\Braket{\pp|\hat p_x\pmm}=-i\Braket{\pp|\partial_x\pmm}
\end{equation}
is purely imaginary, since $\Braket{\pp|\partial_x\pmm}$ is an integral of real
functions. A Hermitian operator on a two-dimensional space whose diagonal
vanishes and whose off-diagonal element is purely imaginary is proportional to
$\sy$, so that
\begin{equation}
-\dot{\xo}\,\hat p_x=\left[-\dot{\xo}\Braket{\pp|\partial_x\pmm}\right]\sy .
\label{eq:boost}
\end{equation}
Lateral displacement of the structure therefore supplies a term of exactly the
form \eqref{eq:Hcd} and of no other form. This is the central result of the
paper. Changes of width and spacing deform the guiding potential in ways that
have no universal generator, and their effect on the coupling must be computed
for each structure; translation is generated by $\hat p_x$ for every structure,
and $\hat p_x$ is purely $\sy$ in any real two-mode basis. Axis bending is
therefore a CD generator by necessity rather than by coincidence.

The same content may be read off the shifted basis
$\phi_a(x,z)=\psi_a\!\left(x-\xo(z);z\right)$, for which the chain rule gives
\begin{equation}
\Braket{\phi_+|\partial_z\phi_-}=\Braket{\pp|\partial_z\pmm}
-\dot{\xo}\Braket{\pp|\partial_x\pmm},
\label{eq:chain}
\end{equation}
that is, as a modification of the intermodal coupling rather than as an added
Hamiltonian term. The two readings are equivalent, and \eqref{eq:chain} makes
contact with the coupling-cancellation language of
Refs.~\cite{Cabanillas2018,Siriani2021}.

\subsection{Closed form for the bending profile}
\label{sec:closed}

Equating the supply \eqref{eq:boost} to the demand \eqref{eq:Hcd}, equivalently
setting \eqref{eq:chain} to zero, determines the axis slope uniquely:
\begin{equation}
\dot{\xo}(z)=\frac{\Braket{\pp|\partial_z\pmm}}{\Braket{\pp|\partial_x\pmm}}
=\frac{-\dot\theta/2}{\Braket{\pp|\partial_x\pmm}},
\qquad
\xo(z)=\int_0^z\dot{\xo}(z')\,dz' .
\label{eq:slope}
\end{equation}
The numerator is the diabatic drive and the denominator the momentum matrix
element that converts axis slope into $\sy$; both are properties of the
unperturbed supermodes at a single cross section, so no trial geometry is
constructed and nothing is re-solved. Where the cancellation condition of
Ref.~\cite{Siriani2021} relates the rates of change of two geometry parameters,
one of which must be probed by finite difference, Eq.~\eqref{eq:slope} closes
the displacement direction analytically and reduces the dimensionality of the
design problem accordingly.

\subsection{Coupled-mode form}
\label{sec:cmt}

Bend-induced perturbations are conventionally expressed through the position
matrix element $\Braket{\pp|x\,\pmm}$ rather than the momentum one. The two are
related by the commutator $[x,\mathcal{H}]=i\hat p_x/m$, which for the
eigenstates of $\mathcal{H}$ gives, exactly,
\begin{equation}
\Braket{\pp|\partial_x\pmm}=k_0n_s\,\dbeta\,\Braket{\pp|x\,\pmm} ,
\label{eq:comm}
\end{equation}
with $\dbeta=\mu_--\mu_+$ the splitting of the eigenvalues of $\mathcal{H}$.
Because $\mathcal{H}$ is an affine function of the scalar Helmholtz operator
$\partial_x^2+k_0^2n^2$, the two share eigenvectors, and in terms of the
Helmholtz propagation constants $\beta_\pm$ the same identity reads
$\Braket{\pp|\partial_x\pmm}=\bar\beta\,(\beta_+-\beta_-)\Braket{\pp|x\,\pmm}$
with $\bar\beta=(\beta_++\beta_-)/2$. The two forms differ by the factor
$\bar\beta/k_0n_s$, which is $1.007$ for the structure of
Sec.~\ref{sec:verify}.
Introducing $\gamma\equiv-2k_0n_s\Braket{\pp|x\pmm}/\dbeta$, the coefficient
through which curvature enters the coupled-mode Hamiltonian of
Ref.~\cite{Lee2026}, Eq.~\eqref{eq:slope} becomes
\begin{equation}
\dot{\xo}(z)=\frac{\dot\theta}{\gamma\,\dbeta^{2}} .
\label{eq:papervars}
\end{equation}
The bending profile is thus fixed by three quantities the coupled-mode
description already carries, namely the rate of change of the mixing angle, the
supermode splitting, and the bend coefficient.

\subsection{Relation to the curvature condition}
\label{sec:frames}

Reference~\cite{Lee2026} derives the bending profile in the
Kramers--Henneberger frame \cite{Longhi2005,Longhi2009}, which is reached from
the comoving frame of Eq.~\eqref{eq:boostH} by the additional tilt phase
$\exp(im\dot{\xo}x')$. The phase removes the boost term and replaces it, up to
a $z$-dependent constant, by a ramp potential linear in $x'$ with slope
$m\ddot{\xo}$. Projected onto the supermodes, the ramp has off-diagonal element
$m\ddot{\xo}X$, with $X\equiv\Braket{\pp|x\,\pmm}$, and diagonal elements
$m\ddot{\xo}\Braket{\psi_\pm|x\,\psi_\pm}$. These are the $\gamma$ and $\alpha$
terms of the bend Hamiltonian $H_b$ of Ref.~\cite{Lee2026}, through which the
bend enters that description in proportion to the curvature. We also note
that in the local basis the diagonal part of the ramp is a phase mismatch
proportional to the curvature, with a magnitude of $m|\ddot{\xo}|d/2$ for
waveguides centered at $\pm d/2$. An axis whose curvature changes sign
therefore results in a sign flip of the phase mismatch, which is the
configuration considered for bent couplers in Ref.~\cite{Wu2022} and for
straight couplers in Ref.~\cite{Huang2020}. The sign flip of the phase mismatch
is thus the boost term of Eq.~\eqref{eq:boostH} viewed in the
Kramers--Henneberger frame.

The slope condition, Eq.~\eqref{eq:slope}, and the curvature condition are
therefore one condition in two pictures. Equation~\eqref{eq:comm} makes the
correspondence explicit on the first-order intermodal amplitude. With
$\Phi(z)=\int_0^z\dbeta\,dz'$, integration by parts gives
\begin{equation}
\int_0^L\dot{\xo}\Braket{\pp|\partial_x\pmm}e^{i\Phi}dz
=-im\left[\dot{\xo}Xe^{i\Phi}\right]_0^L
+im\int_0^L\left(\ddot{\xo}X+\dot{\xo}\dot X\right)e^{i\Phi}dz .
\label{eq:ibp}
\end{equation}
The left-hand side is the amplitude supplied by the boost term in the comoving
frame. On the right, the term in $\ddot{\xo}X$ is the $\gamma$ part of $H_b$;
the boundary term is the tilt phase evaluated at the facets, and vanishes when
$\dot{\xo}(0)=\dot{\xo}(L)=0$; and the term in $\dot{\xo}\dot X$ arises in
Ref.~\cite{Lee2026} from the $z$ dependence of the transformation that returns
the Kramers--Henneberger frame to the waveguide basis. The curvature term is
real and cannot cancel a $\sy$ term pointwise, which is why the cancellation in
that frame is routed through the transformation and leads to an equation for
$\ddot{\xo}$; in the comoving frame the same cancellation reduces to the single
ratio of Eq.~\eqref{eq:slope}.

Two consequences follow. First, $H_b$ is not a correction to be added to
Eq.~\eqref{eq:boostH}: it is the boost term itself, seen from the other frame,
and including both would count the bend twice. Second, the rigidly displaced
supermodes $\phi_\pm$ of Eq.~\eqref{eq:chain} are exact local eigenmodes of the
displaced cross section, so the comoving-frame derivation carries no
mode-deformation error within the paraxial model; the deformation of the bound
modes by the ramp is, inside the two-mode subspace, $H_b$ itself, and outside
it the coupling to other modes discussed in Sec.~\ref{sec:disc}.

\subsection{What the two realizations should and should not share}
\label{sec:should}

Sections~\ref{sec:cd} and \ref{sec:closed} give two devices implementing one
protocol. The gauge realization modifies $\Delta$ and $\Omega$ according to
\eqref{eq:gauge} and leaves the axis straight; the bent realization leaves
$\Delta$ and $\Omega$ unchanged and displaces the axis according to
\eqref{eq:slope}. Both reproduce the dynamics of $H_0+H_{\mathrm{cd}}$, but in
different pictures, related to it by $U_z$ and by the transformation associated
with \eqref{eq:boostH} respectively. This is the sense in which the two
realizations are equivalent up to a frame transformation. At the design
point, each device is described by $H_0+H_{\mathrm{cd}}$ in a frame that
coincides with the laboratory frame at the facets. For the gauge realization
the description is exact within the coupled-mode model, while for the bent
realization, which keeps the spacing of the reference coupler, it holds to
first order in the axis slope.

This has a testable consequence with two halves. The output efficiency, evaluated where both
transformations reduce to the identity, which is to say at the facets, must
agree at the design point. However, its dependence on wavelength and on fabrication
error is expected to be comparable but not identical. This is because the two
devices are different structures away from the design point, so that they
deviate from the protocol in different ways. Quantities evaluated inside the device need not agree, and should not, because
$U_z$ is diagonal in the local basis while the transformation accompanying
\eqref{eq:boostH} is not. Section \ref{sec:equiv} examines both halves.

\section{Numerical verification}
\label{sec:verify}

\subsection{Structure and method}

The reference structure is the polymer mode-evolution coupler of
Ref.~\cite{Lee2026}, on a platform considered for board-level optical
interconnects \cite{Zuo2019,Rasel2022}, with substrate index $n_s=1.5$, index
contrast $\Delta n=0.016$ \cite{Bai2006} and operating wavelength $\lambda=1.55~\mu$m. The widths
$w_1(z)$ and $w_2(z)$ taper linearly from $4.2$ to $2.6~\mu$m and from $2.6$ to
$4.2~\mu$m respectively over the device length $L$, at a fixed center-to-center
spacing $d=5.5~\mu$m. Local modes are obtained from a one-dimensional finite-difference
scalar solver on a $0.02~\mu$m grid using a sub-pixel area-averaged
permittivity, and propagation is computed with a Crank--Nicolson paraxial
scheme \cite{Kawano2001}. The sub-pixel averaging is not a refinement but a requirement: a
staircase index representation advances the taper by $4$~nm per step against a
$20$~nm pixel, and the resulting discretization noise in $\dot\theta$ inflates
the adiabaticity parameter by a factor of ten.

Coupled-mode parameters extracted in this way are $\Omega=2896$--$3107$~m$^{-1}$
and $\Delta=+6270$ to $-6270$~m$^{-1}$, and the untreated coupler transfers
$0.651$, $0.944$ and $0.998$ of the input power at lengths of $500$, $1000$ and
$2200~\mu$m respectively, reproducing the known requirement of more than two
millimeters for near-complete transfer \cite{Chung2023}. Throughout, transfer is quantified by
the supermode fidelity $P_+=|c_+(L)|^2$, the fraction of the output power in the
supermode $\pp$ of the local structure when $\pp$ is launched at the input;
this is the quantity CD driving preserves, and it is evaluated in the frame in
which each device is described, which coincides with the laboratory frame at
the facets.

\subsection{The closed form reproduces the numerical search}
\label{sec:search}

We first compare Eq.~\eqref{eq:slope} against a direct numerical search of the
kind used in Refs.~\cite{Cabanillas2018,Siriani2021}. The discrete search steps along the
propagation direction, tries lateral offsets between consecutive cross
sections, and retains the offset that nulls the adiabaticity parameter
$|\mathcal{A}|/\dbeta$; accumulating these offsets yields $\xo(z)$.
Carried out on the linearly tapered structure above at $L=500~\mu$m, the
search returns
$\dot{\xo}=-0.00643$, $-0.01799$, $-0.03637$, $-0.01797$ and $-0.00643$ at
$z=0$, $125$, $250$, $375$ and $500~\mu$m, against $-0.00645$, $-0.01798$,
$-0.03637$, $-0.01798$ and $-0.00645$ from Eq.~\eqref{eq:slope}. The pointwise
ratio lies between $0.997$ and $1.002$ and the ratio at the peak is $1.0000$.
The commutator identity underlying the coupled-mode form holds to numerical
precision along the device when written with $\bar\beta$ and the Helmholtz
splitting; written with $k_0n_s$ and the same splitting it is $0.67\%$ low, the
factor $\bar\beta/k_0n_s$ noted in Sec.~\ref{sec:cmt}.

\begin{figure}[htbp]
\centering\includegraphics[width=0.75\linewidth]{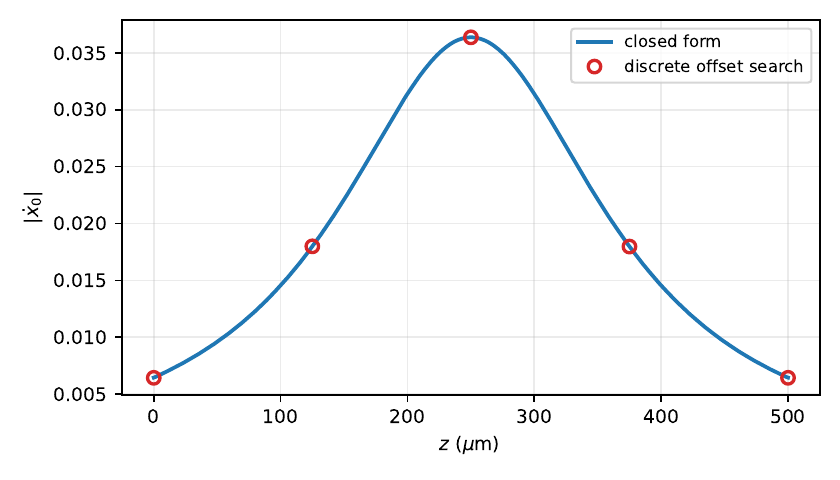}
\caption{\label{fig:design}Axis slope from Eq.~\eqref{eq:slope} compared with a
direct numerical offset search, evaluated at five cross sections of the
linearly tapered reference structure at $L=500~\mu$m. The two agree to $0.3\%$
pointwise.}
\end{figure}

Figure~\ref{fig:design} shows the comparison: the two are indistinguishable on
the scale of the plot, the search points lying on the curve of
Eq.~\eqref{eq:slope} at every cross section sampled. This is the expected
result, since the search locates numerically the same cancellation that
Eq.~\eqref{eq:slope} expresses in closed form, and it confirms that the operator
identity of Sec.~\ref{sec:boost} has not introduced any approximation beyond
those already present in the two-mode description.

The closed form is moreover the more accurate of the two in practice, because
the search degrades when the cross sections are sampled at a realistic spacing:
the extracted peak slope falls to $0.987$ of the continuum value when they are
spaced by $10~\mu$m and to $0.68$ when spaced by $80~\mu$m. Under-resolving the
scan systematically under-bends the device, and nothing in the extracted profile
signals that it has happened.

\subsection{Equivalence of the two realizations}
\label{sec:equiv}

\begin{figure}[htbp]
\centering\includegraphics[width=0.75\linewidth]{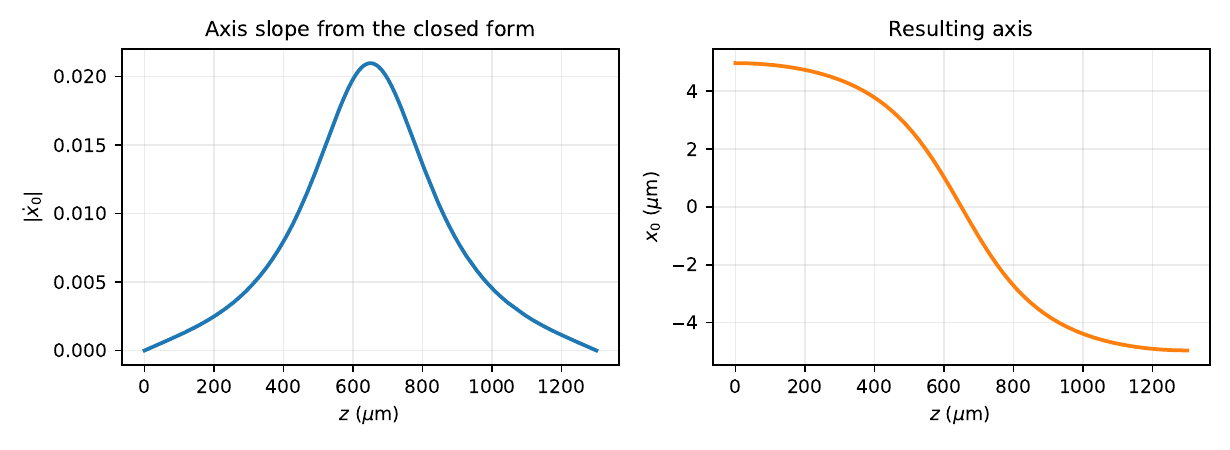}
\caption{\label{fig:design2}Axis slope from Eq.~\eqref{eq:slope} for the
facet-matched taper, and the axis obtained by integrating it. Both the slope and
the gauge angle vanish at the facets, so the two frame transformations reduce to
the identity there.}
\end{figure}

A meaningful comparison of the two devices requires both transformations to
reduce to the identity at the facets, as assumed in
Sec.~\ref{sec:should}. This is a condition on the taper rather than on either
construction: $\varphi=\arctan(\dot\theta/2\Omega)$ and $\dot{\xo}$ of
Eq.~\eqref{eq:slope} are both proportional to $\dot\theta$ at leading order, so
both vanish at the facets if and only if $\dot\theta$ does. A linear width
taper does not satisfy this, and a device built on one carries a facet
discontinuity whose size differs between the two realizations, contaminating any
head-to-head comparison. We therefore take the width profile to follow
$u=3t^{2}-2t^{3}$ with $t=z/L$, for which $\dot u$ vanishes at both ends; the
widths sweep over the same range as before and the cross section is otherwise
unchanged. Because this profile concentrates the variation of $\theta$ near
midstructure, its peak adiabaticity parameter at a given length exceeds that of
the linear taper, and the comparison is made at $L=1300~\mu$m rather than at
the $500~\mu$m of Sec.~\ref{sec:search}; the untreated coupler with this
profile and length transfers $0.886$. The resulting design is shown in
Fig.~\ref{fig:design2}, and has $\varphi(0)=\varphi(L)=0$ and
$\dot{\xo}(0)=\dot{\xo}(L)=0$ to numerical precision.

Both realizations were then constructed from this common $H_0(z)$. The bent
device follows Eq.~\eqref{eq:slope} directly, with a
peak axis slope of $0.021$ and a total lateral excursion of about $10~\mu$m. The
gauge device was obtained by inverting mode-solver tables to realize the dressed
parameters $(\Omega_s,\Delta_s)$ of Eq.~\eqref{eq:gauge}, which requires the
spacing $d$ to vary along the device while the width difference sweeps over
essentially the reference range. Because $\Delta_s$ contains $\ddot\theta$
through $\dot\varphi$, the derivatives of $\theta$ were evaluated from a smooth
fit in the taper coordinate $u$, on which the cross section depends
exclusively, rather than by repeated numerical differentiation of the extracted
$\theta(z)$, which would imprint a point-to-point jitter of up to about $30$~nm
on the width difference.

\begin{figure}[htbp]
\centering\includegraphics[width=\linewidth]{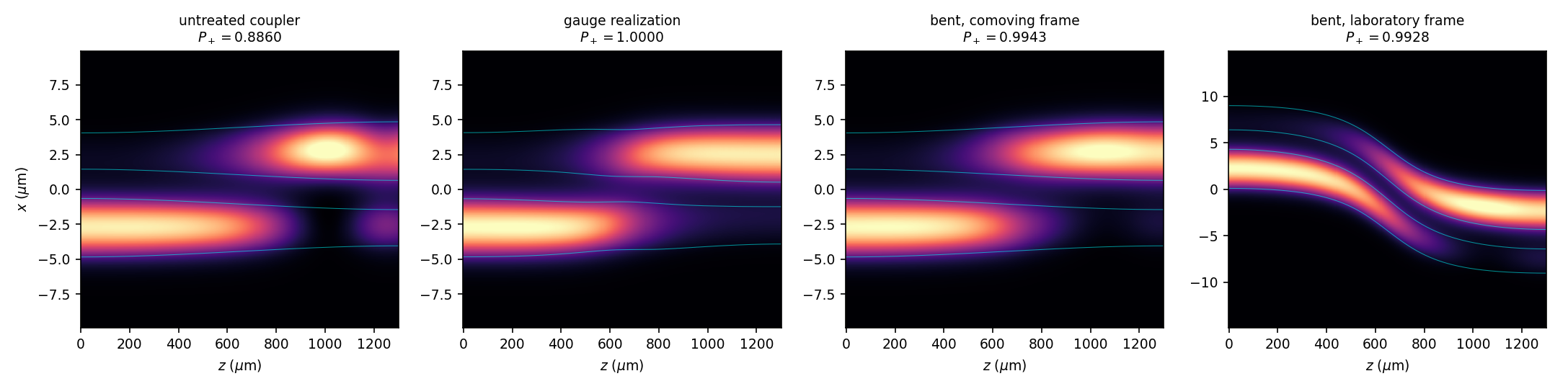}
\caption{\label{fig:fields}Propagated intensity at $L=1300~\mu$m with core
boundaries overlaid. From left: the untreated coupler, the gauge realization,
the bent realization read in the comoving frame $x'=x-\xo(z)$ of
Eq.~\eqref{eq:chain}, in which the structure appears straight, and the same bent
device read in the laboratory frame.}
\end{figure}

Figure~\ref{fig:fields} shows the propagated fields. The untreated coupler
retains $P_+=0.886$; the gauge realization reaches $1.000$ and the
bent realization $0.994$. The third panel reads the bent device in the comoving
frame $x'=x-\xo(z)$ in which Eq.~\eqref{eq:chain} and hence
Eq.~\eqref{eq:slope} are derived, so the structure appears straight and the
displacement is carried by the frame; the fourth panel reads the same device in
the laboratory frame, where the axis excursion is visible directly. They give
$0.994$ and $0.993$, confirming that the two descriptions of one structure agree
at the output. We also note that the remaining $0.006$ of the bent device is left in
$\pmm$ instead of being radiated, with the power outside the two supermodes
below $10^{-5}$ for all three devices.

\begin{figure}[htbp]
\centering\includegraphics[width=\linewidth]{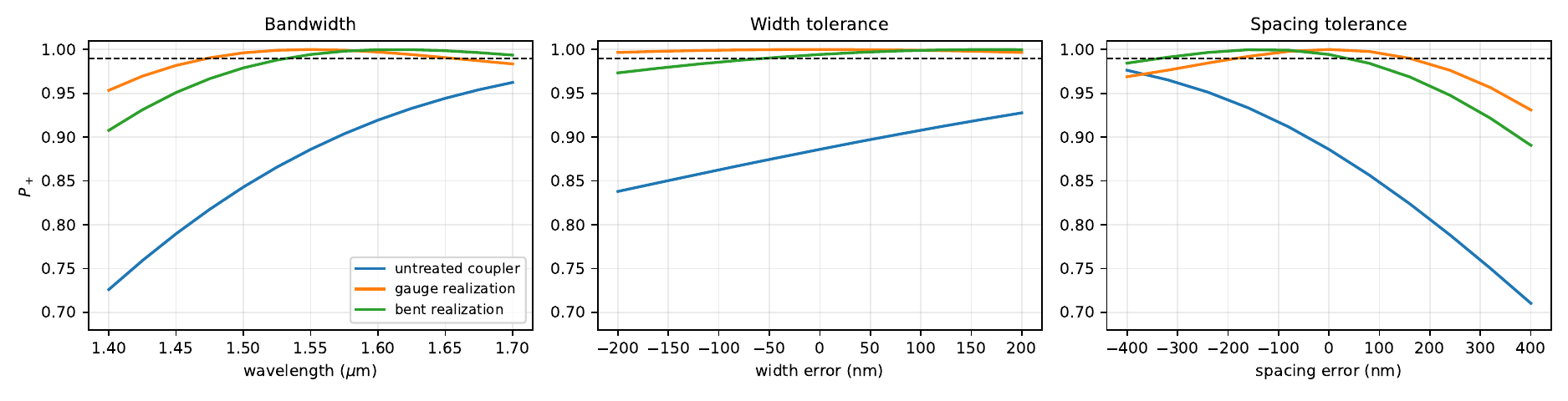}
\caption{\label{fig:sweeps}Supermode fidelity against wavelength and against the
two fabrication error directions, for the untreated coupler and the two
realizations. All values are the output supermode fidelity $P_+$ defined in
Sec.~\ref{sec:verify}, evaluated at the facet, where both frame transformations
reduce to the identity. Dashed line, $P_+=0.99$.}
\end{figure}

The frame-independent observables agree closely, as Fig.~\ref{fig:sweeps} shows.
Across the swept band both realizations exceed $0.99$ over more than $170$~nm
and neither has begun to fall by $1.70~\mu$m, against an untreated coupler that reaches
only $0.962$ at the top of the sweep. A global error in waveguide width is not
the binding constraint for either device, both remaining above $0.97$ across
$\pm200$~nm while the untreated coupler varies by nine points over the same range.
Errors in the spacing $d$ are binding, and the windows above $0.99$ are
comparable in width, roughly $345$~nm for the gauge device and $385$~nm for the
bent one, though offset in center: the bent device tolerates a reduced spacing
better and the gauge device an increased one, consistent with the gauge
realization having already narrowed the spacing to write $\Omega_s$. The two
sets of curves are, however, not expected to coincide, because the equivalence
of Sec.~\ref{sec:should} holds at the design point. For wavelength variations
or fabrication errors, the two structures deviate from the design point
differently. The gauge device deviates through its modulated spacing and width
difference, while the bent device deviates through an axis slope that no longer
satisfies
Eq.~\eqref{eq:slope}.

\begin{figure}[htbp]
\centering\includegraphics[width=0.75\linewidth]{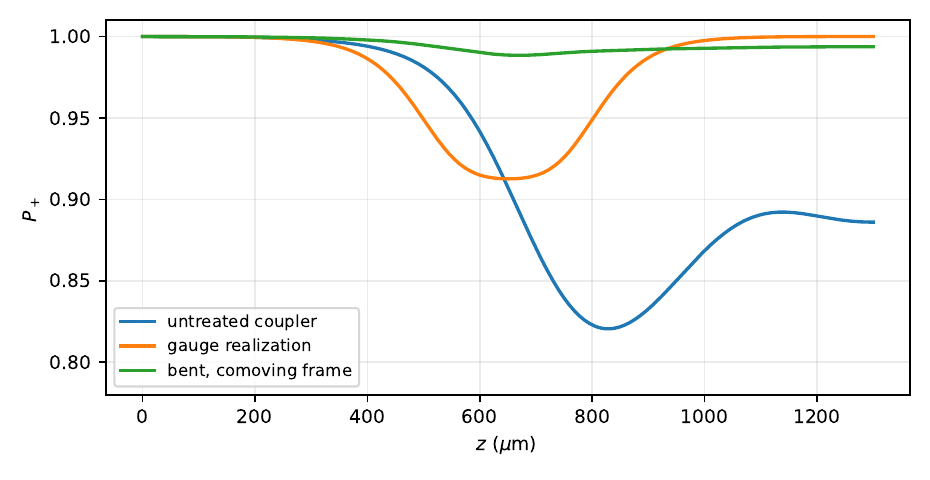}
\caption{\label{fig:power}Supermode fidelity along the device. The bent device is
read in the comoving frame of Eq.~\eqref{eq:chain}, the gauge device in the frame
of Eq.~\eqref{eq:Uz}; both reduce to the laboratory frame at the facets.}
\end{figure}

The trajectories inside the devices differ, and this is the expected signature
rather than a discrepancy. Figure~\ref{fig:power} shows $P_+$ along the device
in the frame in which each realization is described. The untreated coupler
loses population to $\pmm$ at midstructure and recovers only partially. The
gauge realization dips to $0.91$ and returns to unity; the dip is not a loss
but the projection of the dressed state onto the undressed supermode, and its
depth is fixed by the dressing angle of Eq.~\eqref{eq:Uz} through
$1-\sin^2\theta\sin^2(\varphi/2)$. The bent realization, read in the comoving
frame of Eq.~\eqref{eq:chain} in which the boost term appears explicitly in the
Hamiltonian, follows the supermode to within $0.015$ throughout. Read in the
laboratory frame instead, the same device would show a large apparent exchange
between supermodes, since the laboratory supermodes are not the eigenstates of
the comoving Hamiltonian. These excursions are properties of the pictures, not
of the protocol; both vanish at the facets, where the pictures coincide.

\begin{figure}[htbp]
\centering\includegraphics[width=\linewidth]{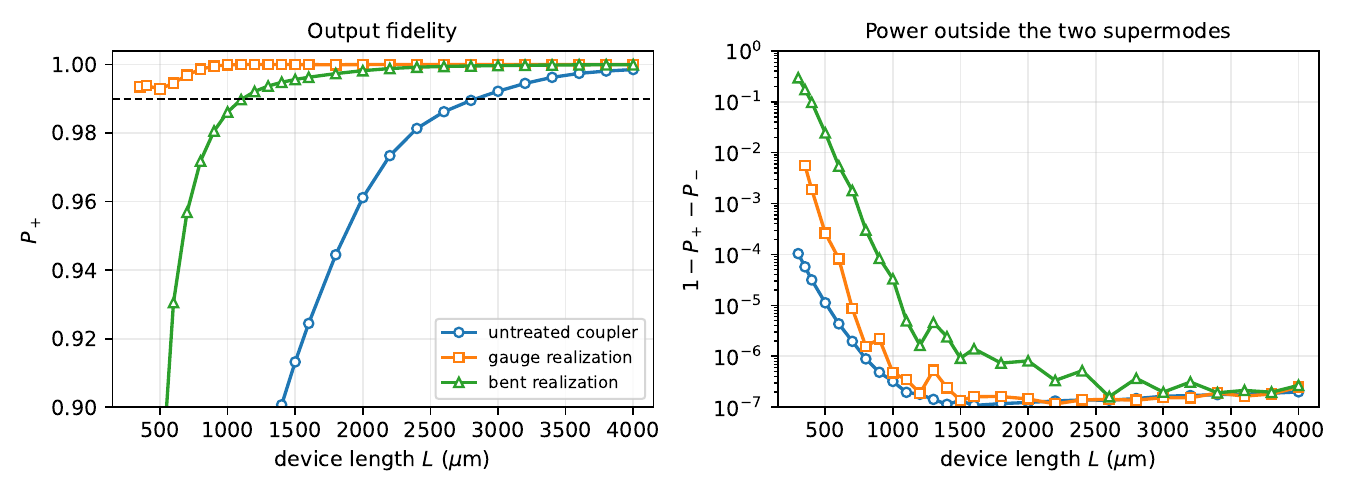}
\caption{\label{fig:length}Output supermode fidelity $P_+$ (left) and power
outside the two supermodes, $1-P_+-P_-$ (right), against device length for the untreated
coupler and the two realizations, each redesigned on the facet-matched taper at
every length. Dashed line, $P_+=0.99$. The gauge realization is shown from
$350~\mu$m, the shortest length at which it was constructed. The floor near
$2\times10^{-7}$ in the right panel is numerical.}
\end{figure}

Finally, we compare the three devices against device length. For each $L$, we
redesign the bent device using Eq.~\eqref{eq:slope} and the gauge device using
Eq.~\eqref{eq:gauge} on the facet-matched taper, and perform beam propagation
simulations with the same numerical settings. In Fig.~\ref{fig:length}, we show
the output supermode fidelity $P_+$ and the power outside the two supermodes,
$1-P_+-P_-$, where $P_-$ is the output power in $\pmm$. From
Fig.~\ref{fig:length}, we can see that the untreated coupler reaches $P_+=0.99$
at $L=2830~\mu$m, while the bent realization reaches it at $1110~\mu$m. The
gauge realization remains above $0.99$ down to $350~\mu$m, where its minimum
gap is reduced from the $2.1~\mu$m of the reference coupler to $0.63~\mu$m.
Shorter gauge designs require a gap below $0.6~\mu$m and were not constructed.
Clearly, the two realizations are limited in different ways, with the bent
device keeping the gap of the reference coupler at every length but radiating
at short lengths. The power outside the two supermodes exceeds $10^{-2}$ when
the peak axis slope exceeds $0.049$ ($L<560~\mu$m), which is about one third of
the acceptance angle $\sqrt{2\Delta n/n_s}=0.146$ of the waveguides. For
$L\geq1000~\mu$m, the power outside the two supermodes is below $10^{-4}$, and the remaining
power of the bent device is left in $\pmm$. This residual is between $4\%$ and $7\%$ of
that of the untreated coupler for device lengths from $1100$ to $4000~\mu$m. On the other hand, the gauge
device is limited by the gap that it requires, and its loss at the shortest
lengths is also due to radiation, which is $5.6\times10^{-3}$ at $350~\mu$m.

\section{Discussion}
\label{sec:disc}

The derivation of Sec.~\ref{sec:theory} is exact within the paraxial
two-level description of the comoving frame, and its limitations are the two places where that
description is incomplete. Both are quantum mechanical in character rather than technological,
and it is worth stating them in those terms.

The first is that $\hat p_x$ does not act within the two-mode subspace. Equation
\eqref{eq:boost} retains the single matrix element $\Braket{\pp|\hat p_x\pmm}$,
but $\hat p_x\pmm$ also has components on the radiation continuum, with matrix
elements $\Braket{k|\hat p_x\pmm}$ that the two-level reduction discards. The
boost term therefore induces transitions not only between the supermodes, where
it cancels the nonadiabatic coupling, but from either supermode into unconfined
modes, where it is a loss channel. In the laboratory frame this is the familiar
tilt loss: a mode propagating at angle $\dot{\xo}$ to its guide carries a
transverse momentum $k_0n_s\dot{\xo}$ that the guided mode does not, and the
mismatch radiates. In the Kramers--Henneberger frame of Sec.~\ref{sec:frames} the same physics
appears as the ramp potential $m\ddot{\xo}x'$ lowering the continuum edge on one
side, so that the
bound supermodes become resonances. The gauge realization has no axis tilt and
no such matrix element, and its radiation at the shortest lengths could be
attributed to the rapid variation of its cross section. Using the length scan
in Fig.~\ref{fig:length}, we can quantify this loss channel. The power outside
the two supermodes of the bent device exceeds $10^{-2}$ when the peak axis slope
exceeds $0.049$, which is about one third of the acceptance angle of the
waveguides. It falls below $10^{-3}$ for slopes under $0.037$ and below
$10^{-4}$ for slopes under $0.030$. The results indicate that a peak slope of
one quarter of the acceptance angle is a practical bound for the bent
realization and for the two-mode description used in deriving
Eq.~\eqref{eq:slope}. At $L=1300~\mu$m, where the peak slope is $0.021$, the
radiated fraction is below $10^{-5}$ for either device, including the extremes
of the sweeps in Fig.~\ref{fig:sweeps}. The insertion loss of the bent device
is $0.03$~dB, which is entirely due to the $0.006$ left in $\pmm$.
The components of $\hat p_x$ outside the two-mode subspace do not only radiate,
but at second order in the axis slope they also act back on the two supermodes,
which is not included in Eq.~\eqref{eq:slope}. This is the origin of the
$0.006$ left in $\pmm$, and the closed form thus cancels the nonadiabatic
coupling to first order in the axis slope.

The second is the paraxial scalar model itself. Within it the bend enters only
through the lateral position of the index profile, and Sec.~\ref{sec:frames}
showed that the comoving-frame treatment then retains the full two-mode content
of the bend, including the curvature term $H_b$ of Ref.~\cite{Lee2026}. What the
model omits are corrections of relative order $\dot{\xo}^{2}$, arising from the
obliquity of the guides with respect to the propagation axis. With peak slopes
of $0.036$ and $0.021$ for the two designs considered here, these are below
$1.3\times10^{-3}$, but they grow as the square of the slope and should be
checked when stronger bending is required, as in compact devices on
high-contrast platforms, where a fully vectorial treatment would in any case be
needed.

The dependence on index contrast can be obtained from a scaling property of
the model. The scalar paraxial equation is invariant under the scaling
$n^2-n_s^2\to s^2(n^2-n_s^2)$, $x\to x/s$ and $z\to z/s^2$, which multiplies
the axis slope and the acceptance angle by the same factor $s$. Within the
model, the bent realization therefore works equally well at higher contrast,
with the device length reduced by a factor of $s^2$. To verify this, we perform
a beam propagation simulation at four times the permittivity contrast with
$L=325~\mu$m. We obtain $P_+=0.886$ for the untreated coupler and $0.995$ for
the bent realization, with a radiated fraction below $10^{-5}$. However, the
axis slope itself increases as $s$, reaching the order of $0.1$ when the
structure is scaled to a silicon nitride core and $0.2$ for a silicon core. In
this case, the corrections of order $\dot{\xo}^{2}$ discussed above and the
vectorial character of the modes are no longer negligible. A fully vectorial
treatment including radiation modes, along the lines of
Ref.~\cite{Siriani2021}, is then required.

The identification of $\hat p_x$ with $\sy$ requires only that the supermodes be
real, which holds for any lossless scalar cross section regardless of its
asymmetry. However, the identification is modified when the modes are complex,
as in a coupler with gain or loss, where the supermodes are biorthogonal. In
this case, both the nonadiabatic coupling and the matrix element of $\hat p_x$ remain antisymmetric
under the unconjugated inner product. Their ratio in Eq.~\eqref{eq:slope} is in
general complex, such that a real axis slope cancels only part of the coupling.
The modes in the Kramers--Henneberger frame carry the tilt phase and are also
complex, which is why the cancellation takes the different form of
Sec.~\ref{sec:frames} in that frame.

The identity can also be extended to more than two modes, since $\hat p_x$ is
imaginary and antisymmetric in any real basis. In this case, a single function
$\xo(z)$ has to cancel more than one nonadiabatic coupling. In a
three-waveguide coupler operated by adiabatic passage \cite{Longhi2009}, the
nonadiabatic term couples the transfer supermode to the other two supermodes
with the same sign. Within nearest-neighbor coupled-mode theory, however,
$\hat p_x$ couples it to them with opposite signs, so that a single bend cannot
cancel both. The CD term of the three-waveguide system is a direct coupling
between the outer waveguides with a $\pi/2$ phase \cite{Chen2010}, which
corresponds to a phase of $\pi/2$ accumulated around the loop formed by the
three waveguides. A rigid bend of a planar coupler cannot provide this term,
because the coupling phases induced by the bend sum to zero around any closed
loop. A possible route is a twisted arrangement of three mutually coupled
waveguides, in which rotation about the propagation axis plays the role of
translation and induces such a phase \cite{Parto2019}, and this will be the
subject of future investigations.

In general, the two realizations follow the same design principle for CD
terms that cannot be provided by the available parameters. Such a term can be
realized when the unitary transformation that generates it corresponds to a physical
operation on the structure and reduces to the identity at its ends
\cite{Ibanez2012}. The gauge realization uses a rotation about the propagation
axis, which is absorbed into the detuning and the coupling, while the bent
realization uses a translation, which is absorbed into the axis position. From
the comparison against device length, we can see the trade-off between the two
realizations. The gauge realization is shorter but reduces the gap to
$0.63~\mu$m, while the bent realization keeps the $2.1~\mu$m gap of the
reference coupler at about three times the length. The results indicate that
axis bending can be used to relax the requirement on the minimum feature size.

\section{Conclusions}

We have identified the transverse momentum $\hat p_x$ as the counter-diabatic
generator of the bent mode-evolution coupler. Rigid lateral displacement of the
structure is generated by $\hat p_x$, whose matrix elements in a real supermode
basis are purely off diagonal and imaginary; displacement therefore supplies a
$\sy$ term and nothing else, which is exactly the form of the CD Hamiltonian
and the form that the detuning and the coupling cannot provide.
Equating the supplied term to the nonadiabatic coupling yields the axis slope
in closed form, Eq.~\eqref{eq:slope}, or equivalently
$\dot{\xo}=\dot\theta/(\gamma\dbeta^{2})$ in coupled-mode variables, and is
the curvature condition of Ref.~\cite{Lee2026} expressed in the comoving rather
than the Kramers--Henneberger frame. The expression agrees with the numerical offset search to $0.3\%$, is more accurate
than a coarsely sampled search, and removes one dimension from the design
problem. Beam propagation simulations of the bent coupler and of the gauge
realization built from the same reference structure give supermode fidelities
of $0.994$ and $1.000$ where the untreated coupler gives $0.886$, with comparable
bandwidth and fabrication tolerance, while the trajectories inside the two
devices differ by the frame transformation that relates them. For a
fidelity of $0.99$, the required device length is $2830~\mu$m for the untreated
coupler, $1110~\mu$m for the bent realization and $350~\mu$m for the gauge
realization, with the gap of the gauge realization reduced to $0.63~\mu$m. The two
realizations are thus one CD protocol implemented in two pictures, and the
bent realization implements it to first order in the axis slope without modulating either the waveguide widths
beyond the original taper or the waveguide spacing. Compared with
Ref.~\cite{Lee2026}, we have provided the physical interpretation of the bend,
a design rule that requires only the axis slope instead of an equation for the
curvature, and the direct comparison of the two realizations.

\begin{backmatter}
\bmsection{Funding}
National Science and Technology Council (114-2221-E-006-049-MY3,
115-2640-E-005-001, 115-2640-E-005-002).

\bmsection{Disclosures}
The authors declare no conflicts of interest.

\bmsection{Data availability}
Data underlying the results presented in this paper are not publicly available
at this time but may be obtained from the authors upon reasonable request.
\end{backmatter}

\bibliography{refs}

\begin{thebibliography}{10}
\newcommand{\enquote}[1]{``#1''}

\bibitem{Dragoman2004}
D.~Dragoman and M.~Dragoman, \emph{Quantum-Classical Analogies} (Springer,
  2004).

\bibitem{Longhi2009}
S.~Longhi, \enquote{Quantum-optical analogies using photonic structures,}
  {\protect\JournalTitle{Laser Photonics Rev.}} \textbf{3}, 243--261 (2009).

\bibitem{Bergmann1998}
K.~Bergmann, H.~Theuer, and B.~W. Shore, \enquote{Coherent population transfer
  among quantum states of atoms and molecules,} {\protect\JournalTitle{Rev.
  Mod. Phys.}} \textbf{70}, 1003--1025 (1998).

\bibitem{Syms1992}
R.~Syms, \enquote{The digital directional coupler: improved design,}
  {\protect\JournalTitle{IEEE Photonics Technol. Lett.}} \textbf{4}, 1135--1138
  (1992).

\bibitem{Xing2013}
J.~Xing, K.~Xiong, H.~Xu \emph{et~al.}, \enquote{Silicon-on-insulator-based
  adiabatic splitter with simultaneous tapering of velocity and coupling,}
  {\protect\JournalTitle{Opt. Lett.}} \textbf{38}, 2221--2223 (2013).

\bibitem{Wang2018}
Y.~Wang, L.~Xu, H.~Yun \emph{et~al.}, \enquote{Polarization-independent
  mode-evolution-based coupler for the silicon-on-insulator platform,}
  {\protect\JournalTitle{IEEE Photonics J.}} \textbf{10}, 1--10 (2018).

\bibitem{Watts2005a}
M.~R. Watts, H.~A. Haus, and E.~P. Ippen, \enquote{Integrated
  mode-evolution-based polarization splitter,} {\protect\JournalTitle{Opt.
  Lett.}} \textbf{30}, 967--969 (2005).

\bibitem{Watts2005b}
M.~R. Watts and H.~A. Haus, \enquote{Integrated mode-evolution-based
  polarization rotators,} {\protect\JournalTitle{Opt. Lett.}} \textbf{30},
  138--140 (2005).

\bibitem{Hung2024}
Y.-J. Hung, C.-H. Chen, H.-C. Chung \emph{et~al.}, \enquote{Compact and
  broadband silicon polarization splitter--rotator using adiabaticity
  engineering,} {\protect\JournalTitle{Opt. Lett.}} \textbf{49}, 1852--1855
  (2024).

\bibitem{Louisell1955}
W.~H. Louisell, \enquote{Analysis of the single tapered mode coupler,}
  {\protect\JournalTitle{Bell Syst. Tech. J.}} \textbf{34}, 853--870 (1955).

\bibitem{Sun2009}
X.~Sun, H.-C. Liu, and A.~Yariv, \enquote{Adiabaticity criterion and the
  shortest adiabatic mode transformer in a coupled-waveguide system,}
  {\protect\JournalTitle{Opt. Lett.}} \textbf{34}, 280--282 (2009).

\bibitem{Ramadan1998}
T.~A. Ramadan and R.~M. Osgood, \enquote{Adiabatic couplers: design rules and
  optimization,} {\protect\JournalTitle{J. Lightwave Technol.}} \textbf{16},
  277--283 (1998).

\bibitem{Syahriar1998}
A.~Syahriar, V.~M. Schneider, and S.~Al-Bader, \enquote{The design of mode
  evolution couplers,} {\protect\JournalTitle{J. Lightwave Technol.}}
  \textbf{16}, 1907--1914 (1998).

\bibitem{Ng2019}
V.~Ng, A.~Tuniz, J.~M. Dawes \emph{et~al.}, \enquote{Insights from a systematic
  study of crosstalk in adiabatic couplers,} {\protect\JournalTitle{OSA
  Continuum}} \textbf{2}, 629--639 (2019).

\bibitem{GueryOdelin2019}
D.~Gu{\'e}ry-Odelin, A.~Ruschhaupt, A.~Kiely, \emph{et~al.}, \enquote{Shortcuts
  to adiabaticity: Concepts, methods, and applications,}
  {\protect\JournalTitle{Rev. Mod. Phys.}} \textbf{91}, 045001 (2019).

\bibitem{Lin2012}
T.~Y. Lin, F.~C. Hsiao, Y.~W. Jhang, \emph{et~al.}, \enquote{Mode conversion
  using optical analogy of shortcut to adiabatic passage in engineered
  multimode waveguides,} {\protect\JournalTitle{Opt. Express}} \textbf{20},
  24085--24092 (2012).

\bibitem{Tseng2012}
S.~Y. Tseng and X.~Chen, \enquote{Engineering of fast mode conversion in
  multimode waveguides,} {\protect\JournalTitle{Opt. Lett.}} \textbf{37},
  5118--5120 (2012).

\bibitem{Tseng2014}
S.-Y. Tseng, \enquote{Robust coupled-waveguide devices using shortcuts to
  adiabaticity,} {\protect\JournalTitle{Opt. Lett.}} \textbf{39}, 6600--6603
  (2014).

\bibitem{Stefanatos2014}
D.~Stefanatos, \enquote{Design of a photonic lattice using shortcuts to
  adiabaticity,} {\protect\JournalTitle{Phys. Rev. A}} \textbf{90}, 023811
  (2014).

\bibitem{MartinezGaraot2017}
S.~Mart{\'\i}nez-Garaot, J.~G. Muga, and S.-Y. Tseng, \enquote{Shortcuts to
  adiabaticity in optical waveguides using fast quasiadiabatic dynamics,}
  {\protect\JournalTitle{Opt. Express}} \textbf{25}, 159--167 (2017).

\bibitem{Chung2019}
H.-C. Chung, S.~Mart{\'\i}nez-Garaot, X.~Chen, \emph{et~al.},
  \enquote{Shortcuts to adiabaticity in optical waveguides,}
  {\protect\JournalTitle{EPL}} \textbf{127}, 34001 (2019).

\bibitem{Wu2020}
Y.-L. Wu, F.-C. Liang, H.-C. Chung \emph{et~al.}, \enquote{Adiabaticity
  engineering in optical waveguides,} {\protect\JournalTitle{Opt. Express}}
  \textbf{28}, 30117--30129 (2020).

\bibitem{Unanyan1997}
R.~Unanyan, L.~Yatsenko, K.~Bergmann \emph{et~al.}, \enquote{Laser-induced
  adiabatic atomic reorientation with control of diabatic losses,}
  {\protect\JournalTitle{Opt. Commun.}} \textbf{139}, 48--54 (1997).

\bibitem{Demirplak2003}
M.~Demirplak and S.~A. Rice, \enquote{Adiabatic population transfer with
  control fields,} {\protect\JournalTitle{J. Phys. Chem. A}} \textbf{107},
  9937--9945 (2003).

\bibitem{Berry2009}
M.~V. Berry, \enquote{Transitionless quantum driving,}
  {\protect\JournalTitle{J. Phys. A: Math. Theor.}} \textbf{42}, 365303 (2009).

\bibitem{Chen2010}
X.~Chen, I.~Lizuain, A.~Ruschhaupt \emph{et~al.}, \enquote{Shortcut to
  adiabatic passage in two- and three-level atoms,}
  {\protect\JournalTitle{Phys. Rev. Lett.}} \textbf{105}, 123003 (2010).

\bibitem{Ibanez2012}
S.~Ib{\'a}{\~n}ez, X.~Chen, E.~Torrontegui, \emph{et~al.}, \enquote{Multiple
  {S}chr{\"o}dinger pictures and dynamics in shortcuts to adiabaticity,}
  {\protect\JournalTitle{Phys. Rev. Lett.}} \textbf{109}, 100403 (2012).

\bibitem{Tseng2013}
S.-Y. Tseng, \enquote{Counterdiabatic mode-evolution based coupled-waveguide
  devices,} {\protect\JournalTitle{Opt. Express}} \textbf{21}, 21224--21235
  (2013).

\bibitem{Paul2015}
K.~Paul and A.~K. Sarma, \enquote{Shortcut to adiabatic passage in a waveguide
  coupler with a complex-hyperbolic-secant scheme,}
  {\protect\JournalTitle{Phys. Rev. A}} \textbf{91}, 053406 (2015).

\bibitem{Okamoto2006}
K.~Okamoto, \enquote{Coupled mode theory,} in \emph{Fundamentals of Optical
  Waveguides,}  (Academic Press, 2006), chap.~4, pp. 159--207, 2nd ed.

\bibitem{Longhi2005}
S.~Longhi, \enquote{Landau--zener dynamics in a curved optical directional
  coupler,} {\protect\JournalTitle{J. Opt. B: Quantum Semiclassical Opt.}}
  \textbf{7}, L9--L12 (2005).

\bibitem{Doerr2005}
C.~Doerr, M.~Cappuzzo, E.~Chen \emph{et~al.}, \enquote{Bending of a planar
  lightwave circuit 2$\times$2 coupler to desensitize it to wavelength,
  polarization, and fabrication changes,} {\protect\JournalTitle{IEEE Photonics
  Technol. Lett.}} \textbf{17}, 1211--1213 (2005).

\bibitem{Morino2014}
H.~Morino, T.~Maruyama, and K.~Iiyama, \enquote{Reduction of wavelength
  dependence of coupling characteristics using {Si} optical waveguide curved
  directional coupler,} {\protect\JournalTitle{J. Lightwave Technol.}}
  \textbf{32}, 2188--2192 (2014).

\bibitem{Tseng2021}
S.-Y. Tseng, \enquote{Robust light coupling in a quadratically bent directional
  coupler,} {\protect\JournalTitle{IEEE Photonics Technol. Lett.}} \textbf{33},
  343--346 (2021).

\bibitem{Wu2022}
B.-T. Wu and S.-Y. Tseng, \enquote{Shortcut to adiabaticity in bent waveguide
  couplers with a sign flip of the phase mismatch,} {\protect\JournalTitle{Opt.
  Express}} \textbf{30}, 25121--25129 (2022).

\bibitem{Chung2023}
H.-C. Chung and S.-Y. Tseng, \enquote{Shortcut to adiabaticity in a bent
  mode-evolution coupler,} {\protect\JournalTitle{Opt. Continuum}} \textbf{2},
  224--231 (2023).

\bibitem{Impens2020}
F.~M.~C. Impens, R.~Duboscq, and D.~Gu{\'e}ry-Odelin, \enquote{Quantum control
  beyond the adiabatic regime in 2d curved matter-wave guides,}
  {\protect\JournalTitle{Phys. Rev. Lett.}} \textbf{124}, 250403 (2020).

\bibitem{Cabanillas2018}
J.~M.~F. Cabanillas and M.~A. Popovi{\'c}, \enquote{Fast adiabatic mode
  evolution based on geometry-induced suppression of nearest-mode crosstalk,}
  in \emph{Conference on Lasers and Electro-Optics (CLEO),}  (OSA, 2018), p.
  STh4A.2.

\bibitem{Siriani2021}
D.~F. Siriani and J.-L. Tambasco, \enquote{Adiabatic guided wave optics -- a
  toolbox of generalized design and optimization methods,}
  {\protect\JournalTitle{Opt. Express}} \textbf{29}, 3243--3257 (2021).

\bibitem{Lee2026}
Y.-K. Lee, Y.~Ban, X.~Chen, and S.-Y. Tseng, \enquote{Counter-diabatic driving
  in a bent mode-evolution coupler,} {\protect\JournalTitle{Opt. Express}}
  \textbf{34}, 1214--1225 (2026).

\bibitem{Yariv1973}
A.~Yariv, \enquote{Coupled-mode theory for guided-wave optics,}
  {\protect\JournalTitle{IEEE J. Quantum Electron.}} \textbf{9}, 919--933
  (1973).

\bibitem{Huang2020}
W.~Huang, L.-K. Ang, and E.~Kyoseva, \enquote{Shortcut to adiabatic light
  transfer in waveguide couplers with a sign flip in the phase mismatch,}
  {\protect\JournalTitle{J. Phys. D: Appl. Phys.}} \textbf{53}, 035104 (2020).

\bibitem{Zuo2019}
H.~Zuo, S.~Yu, T.~Gu \emph{et~al.}, \enquote{Low loss, flexible single-mode
  polymer photonics,} {\protect\JournalTitle{Opt. Express}} \textbf{27},
  11152--11159 (2019).

\bibitem{Rasel2022}
M.~O.~F. Rasel, A.~Yamauchi, and T.~Ishigure, \enquote{Error-free
  three-dimensional multimode crossover graded-index polymer waveguides for
  board-level optical circuitry,} {\protect\JournalTitle{J. Lightwave
  Technol.}} \textbf{40}, 6465--6473 (2022).

\bibitem{Bai2006}
Y.~Bai, Q.~Liu, K.~P. Lor \emph{et~al.}, \enquote{Widely tunable long-period
  waveguide grating couplers,} {\protect\JournalTitle{Opt. Express}}
  \textbf{14}, 12644--12654 (2006).

\bibitem{Kawano2001}
K.~Kawano and T.~Kitoh, \emph{Introduction to Optical Waveguide Analysis:
  Solving Maxwell's Equation and the Schr{\"o}dinger Equation} (Wiley, 2001).

\bibitem{Parto2019}
M.~Parto, H.~Lopez-Aviles, J.~E. Antonio-Lopez, \emph{et~al.},
  \enquote{Observation of twist-induced geometric phases and inhibition of
  optical tunneling via {Aharonov-Bohm} effects,} {\protect\JournalTitle{Sci.
  Adv.}} \textbf{5}, eaau8135 (2019).

\end{thebibliography}

\end{document}